\documentclass[onecolumn,floatfix,superscriptaddress,showpacs,showkeys,nofootinbib,preprint]{revtex4}
\usepackage{epsfig}
\usepackage{amssymb,latexsym,amsmath}
\newcommand{\eq}[1]{\begin{align} #1 \end{align}}

\begin{document}

\title{Particle Spectra in Statistical Models \\
with  Energy and Momentum Conservation}
\author{V.V. Begun}
\affiliation{Bogolyubov Institute for Theoretical Physics, Kiev,
Ukraine}

\author{M. Ga\'zdzicki}
\affiliation{
 Goethe--University, Frankfurt, Germany}
\affiliation{Jan Kochanowski University, Kielce, Poland}

\author{M.I. Gorenstein}
\affiliation{Bogolyubov Institute for Theoretical Physics, Kiev,
Ukraine} \affiliation{Frankfurt Institute for Advanced Studies,
Frankfurt, Germany}

\begin{abstract}
Single particle momentum spectra are calculated  within three
micro-canonical statistical ensembles, namely, with conserved
system energy, system momentum, as well as system energy and
momentum. Deviations from the exponential spectrum of the grand
canonical ensemble are quantified and discussed. For mean particle
multiplicity and temperature, typical for p+p interactions at the
LHC energies, the effect of the conservation laws extends  to
transverse momenta as low as about 3~GeV/c. The results may help
to interpret spectra measured in nuclear collisions at high
energies, in particular, their system size dependence.
\end{abstract}

\pacs{12.40.-y, 12.40.Ee}

\keywords{statistical model, momentum spectra, conservation laws}

\maketitle

\section{ Introduction}

Questions concerning possible phases of strongly interacting matter and
transitions between them have been motivating experimental and theoretical
study of relativistic nuclear collisions for many
years now~\cite{Florkowski_textbook}. Results on collision energy
dependence of hadron production properties in central lead--lead
(Pb+Pb) collisions indicate that a high density phase of strongly
interacting matter, a quark--gluon plasma (QGP), is produced at
an early stage of collisions at energies higher than about
8~GeV (center of mass energy per nucleon--nucleon
pair)~\cite{Gazdzicki:2010iv}.  Signals of the onset of
deconfinement are not observed in proton--proton (p+p)
interactions. This is probably because of a small volume of the
created system. Also, in other cases, when interpreting signatures
of the onset of
deconfinement and/or QGP in nucleus--nucleus (A+A) collisions it is popular
to refer to a comparison with properly normalized data on p+p
interactions at the same collision energy per nucleon.

Statistical models in thermodynamical approximation are, in
general, sufficient to describe mean particle multiplicities in
central Pb+Pb collisions at high energies. This is because the
volume of the created matter is large and therefore an influence
of the material and motional conservation laws can be neglected.
On the other hand, data on p+p interactions are notoriously
difficult to interpret within statistical approaches. This should
be attributed to an importance of the conservation laws and thus
invalidity of thermodynamic models. Instead of the grand canonical
ensemble (GCE), the canonical (CE) or micro-canonical (MCE) ones
should be used.
Of course, this may also impact conclusions from a comparison of
results on Pb+Pb and p+p interactions as well as the study of
system size dependence in A+A collisions.

An influence of material conservation laws on particle yields has
been  studied  within the CE since a long time
\cite{Rafelski:1980gk}. In particular, strangeness~\cite{strCE},
baryon number~\cite{ggg}, and charm~\cite{gk} conservation laws
were considered separately in detail. A complete treatment of the
exact material conservation laws within the CE and MCE
formulations was developed and applied to analyze hadron
yields in elementary collisions~in Refs.~\cite{bec, MCE}. The main
result is that  the density of conserved charge carriers decreases
with decreasing system volume. This so--called CE suppression
becomes significant for a mean multiplicity of conserved charges
of the order of one.

Similar to the CE suppression of particle yields, one may expect
that a shape of single particle momentum spectra changes when
energy and momentum conservation laws are imposed. This conjecture
is addressed quantitatively in our paper in which three
micro-canonical
statistical ensembles, namely, with conserved system energy,
system momentum as well as system energy and momentum, are
considered.

The paper is organized as follows. Partition functions for the
three ensembles are defined and calculated in  Section II. The
corresponding single particle momentum spectra are shown and
discussed in Section III. Summary presented in Section IV closes
the paper.

%
\section{Partition Functions with Conserved Momentum and Energy}
%

For simplicity, a non--interacting gas of  mass-less particles
(without conserved charges) will be studied. Moreover, the
classical Boltzmann approximation, which neglects (small) quantum
effects, will be used. This allows to derive analytical formulas
for single particle spectra in  three micro-canonical
statistical ensembles.
These are ensembles with the fixed volume, $V$, and conserved
system energy $E$ only, system momentum ${\vec P}$ only as well as
$E$ and ${\vec P}$ being conserved together. They are referred to
as $(E,V)$, $(T,{\vec P},V)$, and $(E,{\vec P},V)$ ensembles,
respectively, where $T$ denotes the system temperature. For
comparison, spectra obtained within the GCE, i.e. the $(T,V)$
ensemble, will be used.
The corresponding partition functions read:
 \eq{\label{ZT}
 Z(T,V)
  \;&=\; \sum_{N=0}^{\infty}Z_N(T,V)~=~
  \sum_{N=0}^{\infty}\left[\frac{V}{(2\pi)^3}\right]^N~
  \frac{W_N(T)}{N!}~,\\
 Z(E,V)
\;&=\; \sum_{N=1}^{\infty}Z_N(E,V)~=~
\sum_{N=1}^{\infty}\left[\frac{V}{(2\pi)^3}\right]^N~
  \frac{W_N(E)}{N!}~,\label{ZE}
  \\
 %
  Z(E,\vec{P},V)
\;&=\; \sum_{N=2}^{\infty}Z_N(E,\vec{P},V)~=~
\sum_{N=2}^{\infty}\left[\frac{V}{(2\pi)^3}\right]^N~
  \frac{W_N(E,\vec{P})}{N!}~,\label{ZEP}\\
   Z(T,\vec{P},V)
  \;&=\; \sum_{N=2}^{\infty}Z_N(T,\vec{P},V)~
  =~ \sum_{N=2}^{\infty}\left[\frac{V}{(2\pi)^3}\right]^N~
  \frac{W_N(T,\vec{P})}{N!}~,\label{ZTP}
 }
where
 \eq{\label{WT}
 W_N(T)
 & \;=\; \int d\vec{p}_1\ldots d\vec{p}_N ~
 \exp\left(-~\frac{\sum_{r=1}^N p_r}{T}\right)
        ~=~\left(8\pi~T^3\right)^N~,&N\ge 0~,
    \\
   W_N(E)
 & \;=\; \int d\vec{p}_1\ldots d\vec{p}_N ~
         \delta\left(E~-~\sum_{r=1}^N p_r\right)~,&
         N\ge 1~,\label{WE}
    \\
%
    W_N(E,\vec{P})
 & \;=\; \int d\vec{p}_1\ldots d\vec{p}_N ~
         \delta\left(E~-~\sum_{r=1}^N p_r\right)~
         \delta\left(\vec{P}~-~\sum_{r=1}^N\vec{p}_i\right)~,&
         N \ge 2~,\label{WEP}
    \\
%
   W_N(T,\vec{P})
 & \;=\; \int d\vec{p}_1\ldots d\vec{p}_N ~
 \exp\left(- ~\frac{\sum_{r=1}^N p_r}{T}\right)
  \delta\left(\vec{P}~-~\sum_{r=1}^N\vec{p}_r\right)~,&
   N \ge 2~. \label{WP}
 }
Note, that the minimal possible number of particles in the
ensembles with conserved momentum, Eqs.~(\ref{ZEP}-\ref{ZTP}) and
(\ref{WEP}-\ref{WP}), is $N=2$.

Using  the integral representation of the $\delta$-functions,
 \eq{\label{deltaE}
   \delta\left(E~-~\sum_{r=1}^N p_r\right)~&=~\frac{1}{2\pi}~\int d\alpha\;
         \exp\left(i\alpha\,E
         \;-\;i\alpha\,\sum_{r=1}^Np_r\right)~,\\
\delta\left(\vec{P}~-~\sum_{r=1}^N\vec{p}_r\right)~&=~
\frac{1}{(2\pi)^3}~ \int d^3\vec{\lambda}\;\;
         \exp\left(i\vec{\lambda}\,\vec{P}
   \;-\; i\,\vec{\lambda}\,\sum_{r=1}^N\vec{p}_r\right)~ \label{deltaEP},
 }
one finds:
 \eq{\label{WNE}
 W_N(E) \;&=\; \frac{2^{3N}\pi^N}{(3N-1)!}\;E^{3N-1}\;, \\
 W_N(E,\vec{P})~&
=~\frac{(8\pi~i)^N}{(2\pi)^4}\int d\alpha~
 d^3\vec{\lambda}~\exp\left[i \left(\alpha
 E~+~\vec{\lambda}\vec{P}\right)\right]~\left[\frac{\alpha}
{(\alpha^2~-~\lambda^2)^2}\right]^N~,\label{WNEP}\\
 W_N(T,\vec{P}) \;&=\; \frac{\left(8\pi T^3\right)^N}{(2\pi)^3}
         \int d^3\vec{\lambda}\;\;
         \frac{\exp\left(i\,\vec{\lambda}\,\vec{P}\right)}
         {(1+T^2\lambda^2)^{2N}}~. \label{WNTP}
 }
%
%
%
%
%
Calculating the $\alpha$-- and
$\lambda$--integrals~\cite{Lepore} in Eq.~(\ref{WNEP}) one obtains
%
 \eq{\label{WNEP1}
 W_N(E,\vec{P})
  \;=\;  &\left(\frac{\pi}{2}\right)^{N-1} \frac{(E^2-P^2)^{N-2}}{2P}\;
  \nonumber \\
  &\times~ \sum_{r=0}^NC_N^{\,r}\;
         \frac{(E+P)^{N-r}\;(E-P)^r}{(2N-r-2)!\;(N+r-2)!}\;
         \left[\frac{E+P}{2N-r-1}\;-\;\frac{E-P}{N+r-1}\right]~.
}
%
Performing the summation and rewriting (\ref{WNEP1}) using the
hypergeometric functions one gets:
 \eq{\label{WNEP2}
W_N(E,\vec{P})
  ~& =~  \left(\frac{\pi}{2}\right)^{N-1}\;
  \frac{(E^2-P^2)^{N-2}}{2P}~(E+P)^N
  \nonumber \\
  & \times ~     \bigg[\;\frac{(E+P)}{(N-2)!(2N-1)!}~~
         _2F_1\left(1-2n,\, -n;\, n-1;\,\frac{E-P}{E+P}\right)
 \nonumber \\
 &  -~  \frac{(E-P)}{(N-1)!(2N-2)!}~~
         _2F_1\left(2-2n,\, -n;\, n;\,\frac{E-P}{E+P}\right)
         \;\bigg]~.
  }
Equation~(\ref{WNTP}), after taking the integral, reads:
 \eq{\label{WNTP1}
 W_N(T,\vec{P})
  \;=\;
\frac{1}{(2\pi)^2}\left(\frac{8\pi}{T}\right)^N\sqrt{\pi}
\left(\frac{PT}{2}\right)^{2N
- 3/2}\;
  \frac{K_{2N - 3/2}(P/T)}{(2N-1)!}\;,
  }
where $K_{2N-3/2}(P/T)$ are the modified Bessel functions.
From Eqs.~(\ref{WNEP2}) and (\ref{WNTP1}) follow that
$W_N(E,{\vec P})=W_N(E,P)$ and $W_N(T,{\vec P})=W_N(T,P)$.
Therefore, as it is intuitively expected, the partition functions
(\ref{ZEP}) and (\ref{ZTP}) depend on the absolute value $P$ of
the 3-vector ${\vec P}$, and they are independent of the
${\vec P}$ direction.
For $\vec{P}=0$, from Eqs.~(\ref{WNEP2}) and
(\ref{WNTP1}), one gets:
%
 \eq{\label{WNEP0}
W_N(E,\vec{P}=0)
 & \;=\; \left(\frac{\pi}{2}\right)^{N-1}\;
         \frac{(2N-1)~(4N-4)!}{((2N-1)!)^2~(3N-4)!}~E^{3N-4}
         \;,
 \\
 W_N(T,\vec{P}=0)
 & \;=\; \left(\frac{\pi}{2}\right)^{N-1}\;\frac{(4N - 4)!}{(2N - 1)!(2N -
2)!}\;T^{3N -3}~.\label{WNTP0}
 }
Using Eqs.~(\ref{WT}), (\ref{WNE}), and (\ref{WNEP2}-\ref{WNTP1}),
one obtains the partition functions in the GCE (\ref{ZT}), and the
micro-canonical ensembles (\ref{ZE}-\ref{ZTP}).
Then, the corresponding mean multiplicity is calculated as:
 \eq{\label{Nav}
 \langle N\rangle ~=~ \frac{\sum_{N}N\cdot Z_N}{Z}~.
 }
%

%
%
\section{Single Particle Momentum Spectra}
%
%
The single particle momentum spectrum  in the GCE (\ref{WT})
reads \cite{power}:
 \eq{\label{FT}
  F(p;T)
 \;\equiv\; \frac{1}{\langle N\rangle }\;\frac{dN}{p^2dp}
 \;=\; \frac{V}{2\pi^2\langle N\rangle}\;\exp\left(-\;\frac{p}{T}\right)
 \;=\; \frac{1}{2T^3}\; \exp\left(-\;\frac{p}{T}\right)\;\equiv~F_{Boltz}(p)~.
 }
%
%
The single particle momentum spectra in the micro-canonical
ensembles (\ref{ZE}-\ref{ZTP}) are \cite{power}:
 \eq{\label{FE}
  F(p;E)
 &~=~
 \frac{V}{2\pi^2\langle N\rangle}~\frac{1}{Z(E,V)}~
     \sum_{N=1}^{\infty}~ Z_{N}(E-p,V)\;,\\
 F(p;E,\vec{P}=0)
 &~=~ \frac{V}{2\pi^2\langle N\rangle}~\frac{1}{Z(E,\vec{P}=0,V)}
     ~\sum_{N=2}^{\infty}~ Z_{N}(E-p,p,V)\;, \label{FEP}
 \\
 F(p;T,\vec{P}=0)
&~=~ \frac{V}{2\pi^2\langle
N\rangle}~\frac{\exp\left(-~p/T\right)}{Z(T,\vec{P}=0,V)}
     \sum_{N=2}^{\infty} Z_{N}(T,p,V)\;.\label{FTP}
 }
Relations $Z_N(E-p,{\vec
p},V)=Z_N(E-p,p,V)$  and $Z_N(T,{\vec p},V)=Z_N(T,p,V)$  in
the right--hand--side of Eqs.~(\ref{FEP}) and~(\ref{FTP}),
respectively, were used. Note also that the spectra
(\ref{FT}--\ref{FTP}) satisfy the same normalization condition:
\eq{\label{norm}
\int_0^{\infty}p^2dp~F(p)~=~1~.
}

\begin{figure}[ht!]
 \epsfig{file=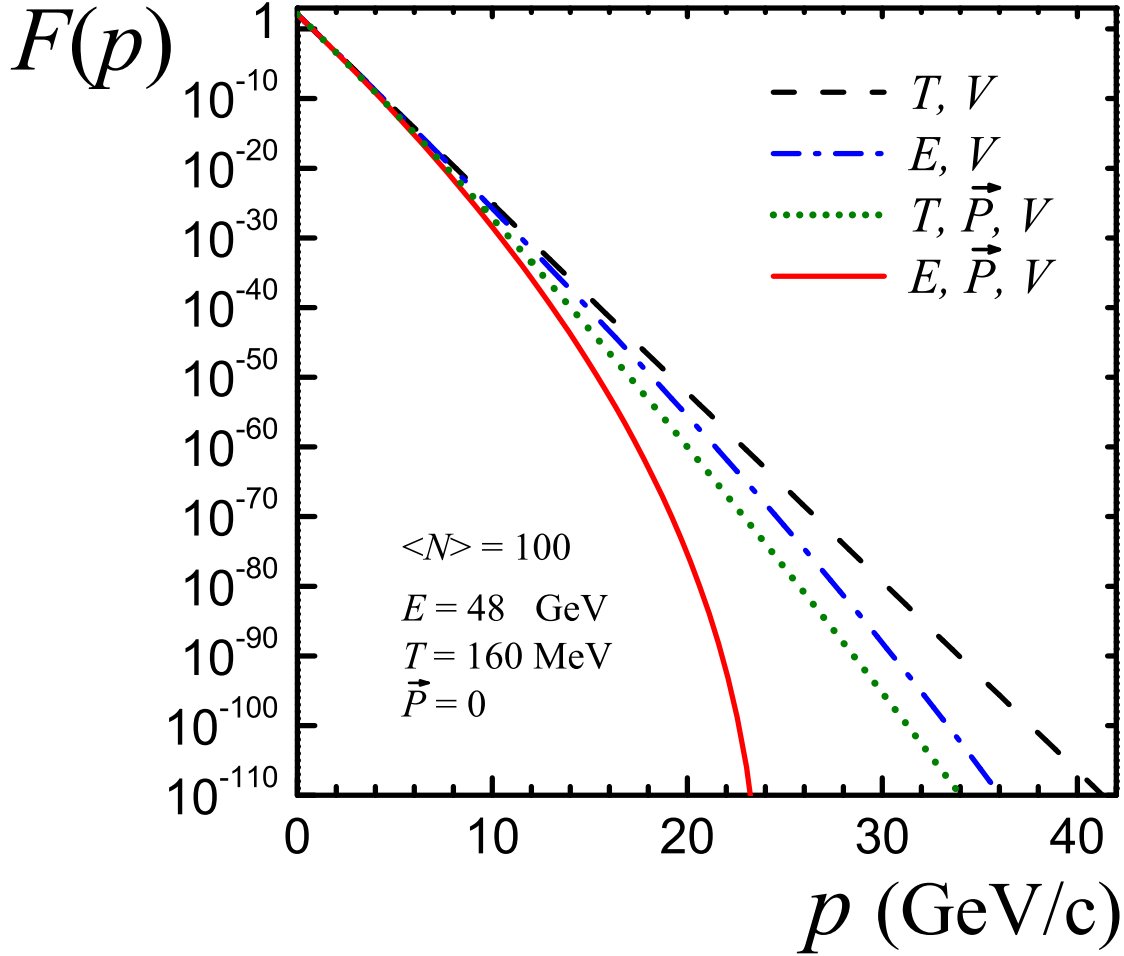,width=0.9\textwidth}\;\;
\caption{Single particle momentum spectra $F(p)$ in the $(T,V)$
(dashed line), $(E,V)$ (dashed--dotted line),  $(E,{\vec
P}=0,V)$ (solid line), and $(T,{\vec P}=0,V)$ (dotted line)
ensembles. } \label{fig1}
\end{figure}

The single particle momentum spectra obtained within the $(T,V)$,
$(E,V)$, $(E,{\vec P},V)$, and $(T,{\vec P},V)$ ensembles  are
shown in  Fig.~\ref{fig1}. They are calculated
for $T = 160$~MeV and $E = 48$~GeV
in the $(T,V)$, $(T,{\vec P},V)$  and $(E,V)$, $(E,{\vec P},V)$
ensembles, respectively. The average multiplicity in the GCE is
selected to be $\langle N\rangle =100$. Note, that the selected mean
multiplicity and temperature are close to those measured  in p+p
interactions at the LHC energies.
The energy in the $(E,V)$ and $(E,{\vec P},V)$ ensembles was set
to be equal to the mean energy in the $(T,V)$ ensemble, i.e.
$E=3T\langle N\rangle$. Finally, the GCE relation $\langle
N\rangle=VT^3/\pi^2$ was used to obtain the volume $V$, which is
used in all ensembles. For these values of $E$, $V$, and $T$ the
average multiplicities (\ref{Nav}) in $(E,V)$, $(E,{\vec P},V)$,
and $(T,{\vec P},V)$ ensembles are then approximately equal to
that in the GCE.

As seen in Fig.~\ref{fig1}, at high momenta the spectra calculated
imposing energy and/or momentum conservation are significantly
below the GCE exponential spectrum. This is expected because  at
the threshold momentum the particle yield has to equal zero,
namely, $F(p;E)\rightarrow 0$ at $p\rightarrow E$, and
$F(p;E,{\vec P}=0)\rightarrow 0$ at $p\rightarrow E/2$.

\begin{figure}[ht!]
 \epsfig{file=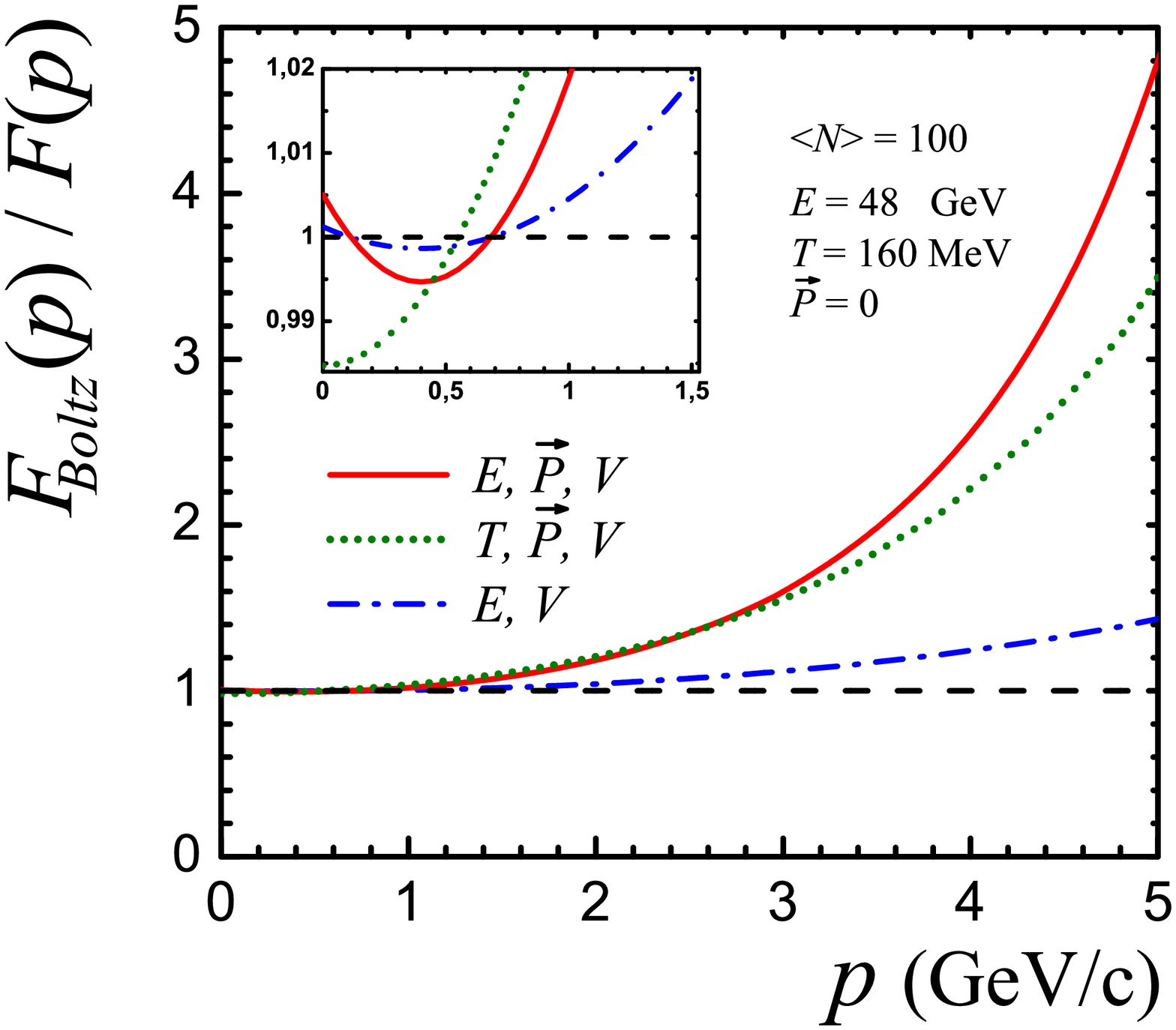,width=0.9\textwidth}
  \caption{
The ratio of $F_{Boltz}(p)/F(p)$, where $F(p)$ equal to
$F(p;E)$ (\ref{FE}), $F(p;E,{\vec P}=0)$ (\ref{FEP}) and
$F(p;T,{\vec P}=0)$  (\ref{FTP}), is shown as a function of
$p$
by the dashed--dotted, solid  and dotted lines, respectively. The
small momentum region, $ p < 1.5 $~GeV/c, is presented in the
inset. Further details are given in the text.
  }
\label{fig2}
\end{figure}

In order to quantify impact of the energy and momentum
conservation at momenta significantly below the threshold one, the
ratio of the spectra in the $(T,{\vec P},V)$, $(E,V)$, and
$(E,{\vec P},V)$ ensembles to the spectrum in the $(T,V)$
ensemble is shown in Fig.~\ref{fig2}.
The suppression of the spectra due to the energy and momentum
conservation is already significant (a factor of about 2) at
momentum $p\cong 3$~GeV/c which is as low as about 12.5\% of the
threshold one. Note, that the ratio is lower than the one by $(1\div
2)$\% at low momenta of several hundred MeV/c as a result of the
suppression of the spectra (\ref{FE}--\ref{FTP}) at higher momenta
and their normalization to unity, Eq.~(\ref{norm}).
\begin{figure}[ht!]
 \epsfig{file=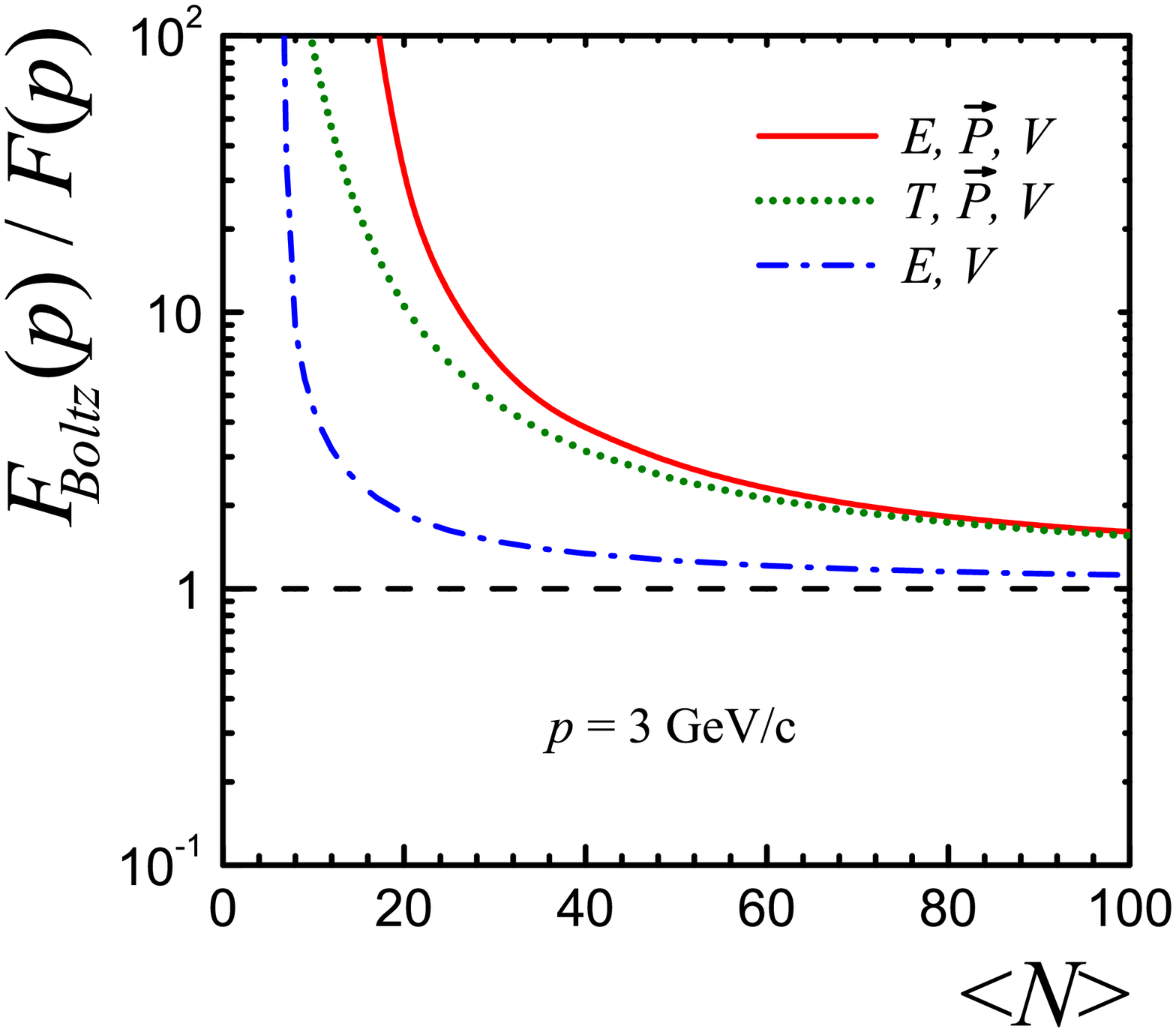,width=0.9\textwidth}\;\;
  \caption{
The ratio of $F_{Boltz}(p)/F(p)$, where $F(p)$ equal to
$F(p;E)$ (\ref{FE}), $F(p;E,{\vec P}=0)$ (\ref{FEP}) and
$F(p;T,{\vec P}=0)$ (\ref{FTP}) is shown as a function of $\langle
N\rangle$ at $ p = 3$~GeV/c by the dashed--dotted, solid  and
dotted lines, respectively. Further details are given in the text.
} \label{fig3}
\end{figure}
Dependence of the ratio $F_{Boltz}(p)/F(p)$  at $p=3$~GeV/c on
mean particle multiplicity is shown in Fig.~\ref{fig3}. The
temperature is fixed as $T=160$~MeV. The energy and volume are
calculated as $E=3T\langle N\rangle$ and $V =\pi^2\langle
N\rangle/T^3$, and they have the same values in all statistical
ensembles. For small statistical systems (at low $\langle
N\rangle$) the effect of the energy--momentum conservation is
strong  and the ratio $F_{Boltz}(p)/F(p)$ is large. The ratio
decreases to unity with increasing $\langle N \rangle$ for all
three ensembles with conserved $E$ and/or ${\vec P}$.  This is
expected, because in the thermodynamical limit an influence of the
energy--momentum conservation on single particle momentum spectra
at any fixed particle momentum should disappear. Note, that in the
$(E,V)$ and $(E,{\vec P},V)$ ensembles,
$F_{Boltz}(p)/F(p)\rightarrow \infty$ at the threshold values of
$\langle N\rangle$ equal to $p/3T$ and $2p/3T$, for $p=E$ and
$p=E/2$, respectively.
\section{Summary}

Single particle momentum spectra are calculated in the system of
mass-less non--interacting Boltzmann particles within three
micro-canonical
statistical ensembles, namely, with the conserved system energy,
system  momentum, as well as system energy and momentum. We
find a strong suppression of the spectra at large momenta in
comparison to the  exponential spectrum of the  grand
canonical (T,V) ensemble. In the $(E,V)$ and $(E,\vec{P}=0,V)$
ensembles the spectra approach zero for momentum approaching its
threshold value, $p\rightarrow E$ and $p\rightarrow E/2$,
respectively. There is no threshold in the $(T,\vec{P}=0,V)$
ensemble, nevertheless the spectrum is also strongly suppressed.
For the mean particle multiplicity and temperature typical for p+p
interactions  at the LHC energy, the suppression of the spectra
due to the energy and momentum conservation is already significant
(a factor of about 2) at $p=3$~GeV/c, i.e. at momenta as low as
about 12.5\% of the threshold momentum for the $(E,\vec{P}=0,V)$
ensemble.

The results of this work are relevant in a study of the system
size and collision energy dependence of transverse momentum
spectra in nuclear collisions at high energies. In particular, an
interpretation of differences between spectra from p+p
interactions and central Pb+Pb collisions should take into account
a possibly strong impact of the energy--momentum conservation.

\begin{acknowledgments} We are thankful to
W.~Greiner  for fruitful discussions.  This work was in part
supported by the Program of Fundamental Research of the Department
of Physics and Astronomy of NAS, Ukraine,  the German
Research Foundation under grant GA 1480/2-1 and
the HICforFAIR grant
20130403.

\end{acknowledgments}
\end{document}